\documentclass{article}
\usepackage{caption}
\usepackage{spconf,amsmath,graphicx}
\usepackage{hyperref}
\usepackage[table,xcdraw]{xcolor}
\usepackage{multirow}
\usepackage{float}                 
\usepackage{tabularx}
\renewcommand{\arraystretch}{1.1}  

\usepackage[table]{xcolor}
\usepackage{adjustbox}

\usepackage{times}
\usepackage{latexsym}
\usepackage{graphicx}
\usepackage{adjustbox}
\usepackage{amsmath}
\usepackage{tabularx}
\usepackage{booktabs} 
\usepackage{array} 
\RequirePackage[T1]{fontenc}
\RequirePackage[utf8]{inputenc}
\RequirePackage{calc}
\RequirePackage{indentfirst}
\RequirePackage{fancyhdr}
\RequirePackage{graphicx,epstopdf}
\RequirePackage{lastpage}
\RequirePackage{ifthen}
\RequirePackage{float}
\RequirePackage{amsmath}
\RequirePackage{amssymb} 
\RequirePackage{mathpazo}
\RequirePackage{booktabs} 
\RequirePackage{titlesec}
\RequirePackage{etoolbox} 
\RequirePackage{tabto} 
\RequirePackage{xcolor, colortbl} 
\RequirePackage{soul} 

\RequirePackage{multirow}
\RequirePackage{microtype} 
\RequirePackage{tikz} 
\RequirePackage{refcount} 
\RequirePackage{changepage} 
\RequirePackage{attrib} 
\RequirePackage{upgreek} 
\RequirePackage{array} 
\RequirePackage{tabularx}
\RequirePackage{pbox} 
\RequirePackage{ragged2e} 
\RequirePackage[]{tocloft} 
\RequirePackage{marginnote} 
\reversemarginpar 
\RequirePackage{marginfix} 
\RequirePackage{enotez} 
\RequirePackage{xstring} 
\usepackage[T1]{fontenc}

\usepackage[utf8]{inputenc}

\usepackage{microtype}

\usepackage{inconsolata}

\title{Comparative Analysis of Modality Fusion Approaches for
Audio-Visual Person Identification and Verification}

\name{Aref Farhadipour\thanks{Correspondence: \texttt{aref.farhadipour@uzh.ch}},Masoumeh Chapariniya, Teodora Vukovic, Volker Dellwo
}

\address{
Department of Computational Linguistics, University of Zurich, Switzerland \\
\texttt{masoumeh.chapariniya@uzh.ch} \\
\texttt{teodora.vukovic2@uzh.ch} \\
\texttt{volker.dellwo@uzh.ch}}

\begin{document}
\maketitle
\begin{abstract}
Multimodal learning involves integrating information from various modalities to enhance learning and comprehension. We compare three modality fusion strategies in person identification and verification by processing two modalities: voice and face.
In this paper, a one-dimensional convolutional neural network is employed for x-vector extraction from voice, while the pre-trained VGGFace2 network and transfer learning are utilized for face modality. In addition, gammatonegram is used as speech representation in engagement with the Darknet19 pre-trained network. The proposed systems are evaluated using the K-fold cross-validation technique on the 118 speakers of the test set of the VoxCeleb2 dataset. The comparative evaluations are done for single-modality and three proposed multimodal strategies in equal situations. Results demonstrate that the feature fusion strategy of gammatonegram and facial features achieves the highest performance, with an accuracy of 98.37\% in the person identification task. However, concatenating facial features with the x-vector reaches 0.62\% for EER in verification tasks.
\end{abstract}

\section{Introduction}

Biometric modalities encompass distinct static physiological traits that remain consistent within the human body, like fingerprints, as well as dynamic behavioral traits that are unique characteristics displayed in response to interactions with the environment, such as gait. However, some modalities can be a combination of both static and dynamic, such as speech \cite{minaee2023biometrics}. In the real world, the human brain simultaneously processes multiple modalities to recognize the identity of each person. Although the exact mechanism of multimodal processing in the human brain remains unclear, the human mind can effortlessly identify individuals based on their faces and voices with minimal errors \cite{perrodin2015brain}.
Voice and face have garnered significant attention in the development of automatic identity recognition systems \cite{minaee2023biometrics, farhadipour2023facial, farhadipour2024analysis}.
While both modalities can change over time or be vulnerable to spoof attacks, combining and analyzing mixed forms of these two modalities can potentially increase the uniqueness of biometric features for each person in identity recognition. In human-machine interaction, identity recognition plays a crucial role. It has been applied in various tasks such as access control, automatic monitoring of older individuals, rehabilitation programs for people with physical and mental disabilities, etc. 
Automatic identity recognition can be categorized into two main tasks: person verification and person identification. Person verification involves authenticating a claimed identity. This typically involves a two-class classification, comparing the claimed identity with unique specific and universal background models. In other words, the system determines whether to accept or reject the claimed identity. On the other hand, person identification refers to the process of identifying an identity in a multi-class classification scenario, where the system needs to determine the person's identity from a large pool of individuals who have been previously trained. While there has been significant research on developing multimodal person verification systems \cite{shah2023speaker}, there is still limited work done in the context of multimodal person identification.

In recent years, deep learning has emerged as an effective approach to pattern recognition, allowing for the modeling of complex functions. This approach enables multimodal signal processing and can be implemented through different tasks. In this work, we utilized deep learning approaches in identity recognition tasks with three different strategies. The first strategy is sensor fusion, where raw information, such as sound and image data, are directly integrated into a classifier. This approach combines the information from different sources at the input level without front-end processing. 
The second strategy is feature fusion, which involves separately extracting low-dimensional features from each modality and then combining them to feed into a classifier. This approach allows for the extraction of modality-specific features before integration. The third strategy is score fusion, where separate classifiers are designed for each modality, and the final scores from each classifier are merged and fed into a decision-making system.

In this work, it is assumed that the person's voice and face information are accessible simultaneously. We trained two separate systems to learn and recognize each modality individually for person identification and verification. Furthermore, three multimodal strategies have been proposed, including all three discussed modes: sensor-level fusion, feature-level fusion, and score-level fusion in identification tasks. Finally, the best fusion mode that is explored in the identification task is utilized in the verification scenario.

To achieve these objectives, the pre-trained VGGFace2 \cite{cao2018vggface2} is utilized for the visual modality. However, for the voice modality, we proposed two methods consisting of gammatonegram representation \cite{farhadi2023gamma} and x-vector \cite{snyder2018x}. This approach results in the creation of the proposed FaceNet and VoiceNet models for each modality from the VoxCeleb2 dataset. 

In the feature and score fusion modes, we mixed facial features with gammatonegram and x-vector separately, and a softmax layer was employed for multimodal learning by combining the information from both modalities. However, in score fusion mode, because we chose a two-dimensional space for mixing the modalities, we just concatenated facial features with gammatonegram representation. The proposed systems are trained and evaluated using the test section of the VoxCeleb2 dataset, which consists of 118 speakers. A K-fold cross-validation approach is utilized to ensure robustness and reliability.

The rest of the paper is organized as follows. Section \ref{sec:RW} presents an overview of related works. In Section \ref{sec:proposedML}, the proposed strategies for multimodal learning in person identification and verification are described in detail. Section \ref{sec:EvalSet} focuses on the evaluation setup. Experimental results are reported in section \ref{sec:Results}. Discussing the results and comparison with previous works are done in section \ref{sec:dis}. Finally, section \ref{sec:cocl} concludes the work and discusses some ideas for future trends.

\section{Related Works}
\label{sec:RW}

Multimodal learning systems in identity identification integrated different modalities, including the fusion of fingerprint and DNA \cite{venkata2022hybrid}, face and gait \cite{prakash2023multimodal}, and face, palmprint and iris \cite{aldjia2021sensor}. In integrating face and speech, the main works focused on speaker verification \cite{wang2022multi}, and few works are accomplished in speaker identification. It should be noted that the existence of the annual VoxSRC challenge had a great impact on this tendency \cite{huh2023voxsrc}. In this part, we report some of the previous works in multimodal speaker identification.

In the paper \cite{chung2018voxceleb2}, an EER of 4.42\% was reported on the Voxceleb1 dataset in a speaker verification task. The same authors in another article \cite{nagrani2020voxceleb} reported 2.95\% as EER on the Voxceleb1 dataset. Authors in paper \cite{moufidi2023attention} utilized a residual neural network to encode depth videos, while a time delay neural network architecture was used to encode voice signals. To evaluate the performance, 1,000 random speakers from the VoxCeleb2 dataset were selected. In that work, the accuracy of the three systems based on voice, depth of the lip regions, and overall fusion reached 56.03\%, 17.49\%, and 64.11\%, respectively.

Alam et al., \cite{alam2015confidence} introduced a confidence-based score fusion framework for audio-visual biometric identification. They proposed the confidence-ratio approach. The researchers utilized a database consisting of 43 individuals. The results indicated that, in the best case with clean data, an accuracy of 90.45\% was achieved. Paper \cite{asadpour2011audio} employed a combination of audio and video features by implementing a hidden Markov model. 

The study \cite{shah2023speaker} suggested a two-branch network to extract facial and voice signal features, using a support vector machine to classify speakers based on single and multi-domain features. Their method got an accuracy of 97.2 \% for Speaker identification performance on VoxCeleb1. In \cite{sari2021multi}, a multi-view model with a shared classifier to map audio and video into a unified space was introduced. The unimodal and audio-visual fusion approaches achieved an equal error rate of 1.6\% on the VoxCeleb2 dataset in the person verification.  

Qian et al.,\cite{qian2021audio} introduced three types of audio-visual deep neural networks: feature level, embedding level, and embedding level combination with joint learning. The VoxCeleb2 test trial list experiments demonstrated equal error rates of 5.08\% and 2.89\% for visual and audio modality systems, respectively. Tao et al., \cite{tao2020audio} employed an audio-visual cross-modal discrimination network for speaker recognition. The best setting of the system gained an accuracy of 86.12\% on VoxCeleb2 for speaker identification in score-level fusion.

The paper \cite{stefanidi2020application} introduced an approach to person identification using CNNs on the VoxCeleb1 audiovisual database. The outcomes achieved an 86.97\% top-5 accuracy. Cai et al., \cite{cai2022incorporating} presented a self-supervised learning framework for speaker recognition, which combined clustering and deep representation learning. With multi-modal training data, their framework obtained an accuracy of 77.60\% on the VoxCeleb2.

\section{Multimodal Learning Strategies}
\label{sec:proposedML}

\begin{figure*}[t]
  \centering
  \includegraphics[width=0.90\textwidth]{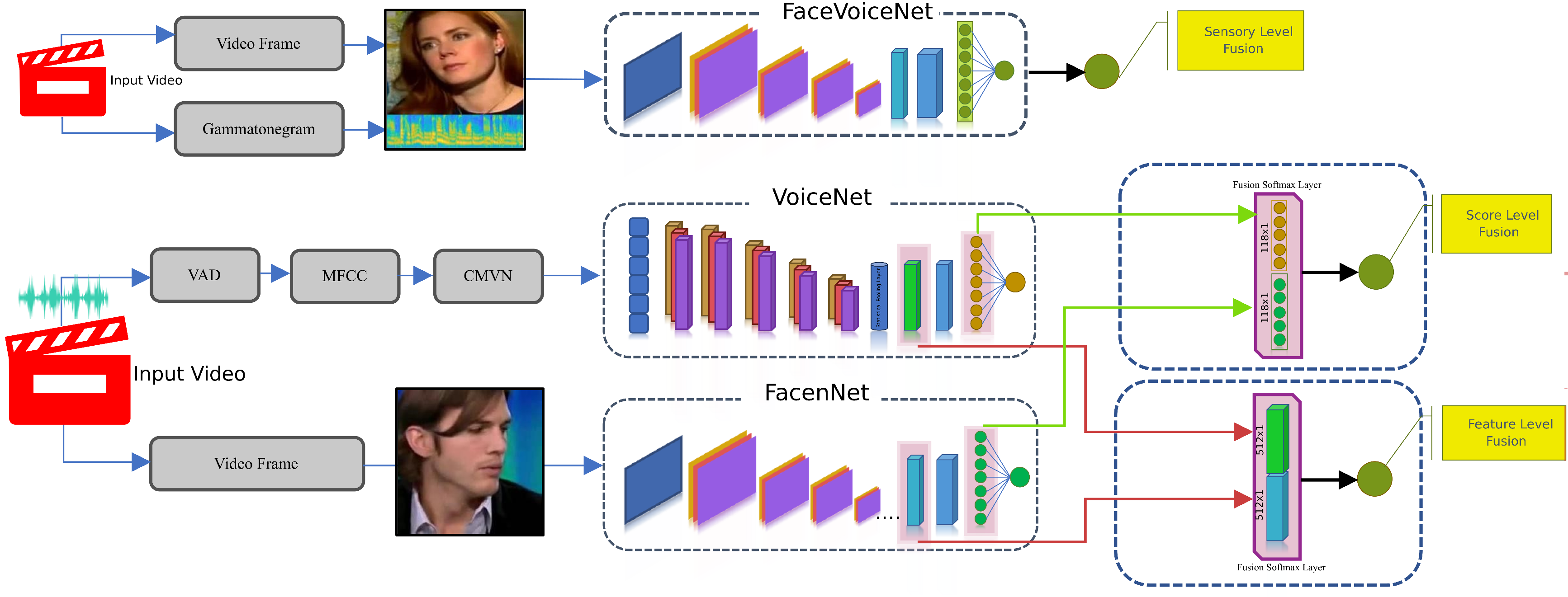}
  \caption{The block diagram of the proposed single-modality systems consists of VoiceNet, FaceNet, and three multimodal systems with different fusion strategies}
  \label{fig:f1}
\end{figure*}

This work uses two single-modality systems to analyze each modality before exploring multimodal learning strategies. For the first system, VoiceNet, we examined two methods. The first one is the x-vector, which is a one-dimensional CNN with five convolutional layers, a statistical pooling layer \cite{wang2021revisiting}, and three fully connected layers that focus on identity identification from voice modality. The architecture of this neural network is depicted in Figure \ref{fig:f1}. The second one is the gammatonegram, which represents the utterances as an image and fine-tunes the Darknet19 \cite{redmon2017yolo9000} for learning the features of each speaker and classification task.

As shown in figure \ref{fig:f1}, the person's speech is extracted from the video file, and silence is removed using a Voice Activity Detector (VAD)
\cite{gian2009method}. The audio files are then transformed into MFCC (Mel-frequency cepstral coefficients) features, and the mismatch between training and test utterances is reduced using the CMVN (Cepstral Mean and Variance Normalization) technique \cite{prasad2013improved} and serves as input for VoiceNet. We made another VoiceNet using gammatonegram separately. The final layer of VoiceNet consists of 118 neurons, representing the number of speakers. The outputs of VoiceNets present the performance of the person identification systems based on voice modality.

In addition to VoiceNet, the FaceNet is proposed for face recognition, as shown in Figure \ref{fig:f1}. FaceNet is built using a pre-trained CNN known as VGGFace2, which has been trained on more than 3.3 million face images \cite{cao2018vggface2}. The transfer learning technique is employed, where the final layers of the network are replaced with new layers to learn the specific information of the new individuals.

The output of FaceNet represents the result of person identification using the face modality. To ensure comparability across different multimodal learning scenarios, an overall audio file and a single frame of the face are extracted from each video file. Since each video contains only one person, the 25th frame is extracted as the face modality.

Figure \ref{fig:f1} illustrates the three proposed strategies for multimodal learning: sensor fusion, feature fusion, and score fusion. This study uses FaceNet and VoiceNets as the foundation for feature and score fusion systems. However, We employ a separate network with mixed modalities as input for the sensor fusion mode. In the subsequent sections, we comprehensively explain each multimodal learning strategy.
\subsection{Sensor Level Fusion}
\label{ssec:sensorF}
Since images are two-dimensional and sounds are one-dimensional signals in this multimodal learning strategy, we aim to combine them into a shared data space. We chose the gammatonegram visualization method \cite{Farhadipour2014gammatonegram} to represent audio files. Sensor-level multimodal learning is performed on the VGGFace2 network and using transfer learning. As depicted in the sensor level fusion system in Figure \ref{fig:f1}, we create an integrated image to serve as input for the FaceVoiceNet. This input image incorporates both facial and voice information. The reasons for choosing voice as the modality to be mapped to a new space were influenced by two factors: the availability of a powerful pre-trained network for face recognition and the common practice of representing voice in the form of an image using the gammatonegram.
\subsection{Feature Level Fusion}
\label{ssec:featureF}
In the feature fusion mode, as depicted in Figure \ref{fig:f1}, a softmax layer is employed for multimodal learning. In this mode, VoiceNet is utilized to extract the x-vector, resulting in a feature vector of speakers with dimensions of 512x1. Similarly, from the FaceNet, the activation values of the last ReLU layer are extracted as the facial feature embedding, which is a 512x1 vector. These two vectors are concatenated and construct a bimodal vector with a size of 1024x1. In another scenario, we replace gammatonegram and Darknet with the x-vector. In this situation, we extract a 118x1 vector from a middle layer of Darknet19 as a feature vector to concatenate with facial features and create a 630x1 dimension multimodal feature vector.

\subsection{Score Level Fusion}
\label{ssec:scoreF}
The score fusion strategy aims to reach a function that accurately predicts the correct speaker ID by using the score vectors from the basis networks. In other words, our proposed system tries to estimate the \textit{f} in eq. \eqref{eq:e1} to make \textit{D}, which is the ground truth in the training process and the final decision in the testing phase.
\begin{equation} \label{eq:e1}
 D=f(x , y)
\end{equation}
In this equation, \textit{x} and \textit{y} are the score vectors of VoiceNets and FaceNet, respectively. We used the softmax layer for modeling \textit{f} based on its high modeling ability. Each input vector has 118 components, resulting in a bimodal vector with a dimension of 236x1.

\section{Evaluation Setup}
\label{sec:EvalSet}
In this work, the test part of the VoxCeleb2 dataset is used for evaluation. This part consists of 118 speakers and is extracted from YouTube videos in real-world conditions, including various types of noise such as laughter, cross-talk, channel effects, music, and some other environmental sounds. We used this subset just to compare the performance of different strategies proposed in this work in a common situation. The test section of VoxCeleb2 comprises a total of 4,911 unique video files and 36,237 utterances extracted from these videos \cite{chung2018voxceleb2}. To evaluate the performance of the systems, the K-fold cross-validation technique is employed, with a value of K=3 based on a traditional strategy to use 70\% of data for training and the rest for testing. This means that the evaluation is repeated three times, and in each iteration, two folds of the dataset are used for training, while the remaining fold is employed for testing. Under these conditions, approximately 24,000 bimodal utterances are used for training, while around 12,000 bimodal utterances are used for testing. 

VoiceNet was trained in 8 epochs, with 128 batch sizes, and the learning rate was adjusted from 1e-3 to 1e-5. On the other hand, the VGGFace2 is trained in 15 epochs with batch size 32 and a constant learning rate of 1e-3. For fine-tuning the Darknet19, we set the learning rate to 1e-4 with 15 epochs. Cross-entropy was used as a loss function in the training process of CNNs and softmax layers. Moreover, stochastic gradient descent with momentum was utilized as the optimizer. The training process is stopped for multimodal learning using the softmax layer when the minimum gradient criteria are satisfied. In this work, it occurred around the 40th to 50th epoch. Two data augmentation techniques were used to avoid overfitting and making a general system: rotation operators within the range of [-20, 20] degrees and vertical and horizontal translation by a distance of [5, 5] pixels.  

To evaluate proposed identification systems, there exist several metrics that can be utilized to evaluate the efficiency of a multi-class pattern recognition system. The system's effectiveness can be demonstrated from various perspectives by carefully selecting the appropriate metrics that enable comparisons with other works. The proposed evaluation parameters encompass precision, specificity, sensitivity, accuracy, and F1 score.  The utilization of a confusion matrix allows for a visual representation of these parameters \cite{lever2016class}. Verification systems have encountered two types of errors: false acceptance and false rejection. False acceptance refers to the claims that were accepted inaccurately. However, false rejection deals with the identity that is rejected incorrectly. Equational Error Rate (EER) is the optimum point at which these two errors are equal.

\section{Experimental Results}
\label{sec:Results}
Our main goal is to compare different strategies in modality fusion. In this part, we report experimental results in two scenarios. Person identification results are reported to compare the performance of different fusion modes in multi-class classification scenarios in equal situations. Based on the results of the identification task, the best fusion mode is evaluated in the verification scenario as a two-class classification besides single modality modes.

\begin{table*}

\small
\centering
\adjustbox{width=15cm}{
\begin{tabular}{
>{\columncolor[HTML]{FFFFFF}}l |
>{\columncolor[HTML]{FFFFFF}}c
>{\columncolor[HTML]{FFFFFF}}c
>{\columncolor[HTML]{FFFFFF}}c
>{\columncolor[HTML]{FFFFFF}}c |
>{\columncolor[HTML]{FFFFFF}}c
>{\columncolor[HTML]{FFFFFF}}c
>{\columncolor[HTML]{FFFFFF}}c
>{\columncolor[HTML]{FFFFFF}}c |
>{\columncolor[HTML]{FFFFFF}}c
>{\columncolor[HTML]{FFFFFF}}c
>{\columncolor[HTML]{FFFFFF}}c
>{\columncolor[HTML]{FFFFFF}}c |}
\cellcolor[HTML]{FFFFFF}                          & \multicolumn{4}{c|}{\cellcolor[HTML]{D9D9D9}Face Identification}                                                                                                                                                       & \multicolumn{4}{c|}{\cellcolor[HTML]{BFBFBF}x-Vector Speaker Identification}                                                                                                                                                        & \multicolumn{4}{c|}{\cellcolor[HTML]{A6A6A6}Gamma. Speaker Identification}                                                                                                                                                      \\ \cline{2-13}
\multirow{-2}{*}{\cellcolor[HTML]{FFFFFF}Metrics} & \multicolumn{1}{l}{\cellcolor[HTML]{FFFFFF}Fold 1} & \multicolumn{1}{l}{\cellcolor[HTML]{FFFFFF}Fold 2} & \multicolumn{1}{l}{\cellcolor[HTML]{FFFFFF}Fold 3} & \multicolumn{1}{l|}{\cellcolor[HTML]{D9D9D9}Avg.} & \multicolumn{1}{l}{\cellcolor[HTML]{FFFFFF}Fold 1} & \multicolumn{1}{l}{\cellcolor[HTML]{FFFFFF}Fold 2} & \multicolumn{1}{l}{\cellcolor[HTML]{FFFFFF}Fold 3} & \multicolumn{1}{l|}{\cellcolor[HTML]{BFBFBF}Avg.} & \multicolumn{1}{l}{\cellcolor[HTML]{FFFFFF}Fold 1} & \multicolumn{1}{l}{\cellcolor[HTML]{FFFFFF}Fold 2} & \multicolumn{1}{l}{\cellcolor[HTML]{FFFFFF}Fold 3} & \multicolumn{1}{l|}{\cellcolor[HTML]{A6A6A6}Avg.} \\ \hline
Precision                                         & 0.97                                               & 0.97                                               & 0.95                                               & \textbf{0.96}                                     & 0.68                                               & 0.77                                               & 0.74                                               & \textbf{0.73}                                     & 0.60                                               & 0.64                                               & 0.60                                               & \textbf{0.61}                                     \\
Sensitivity                                       & 0.97                                               & 0.97                                               & 0.95                                               & \textbf{0.96}                                     & 0.68                                               & 0.77                                               & 0.74                                               & \textbf{0.73}                                     & 0.60                                               & 0.64                                               & 0.60                                               & \textbf{0.61}                                     \\
Specificity                                       & 1.00                                               & 1.00                                               & 1.00                                               & \textbf{1.00}                                     & 1.00                                               & 1.00                                               & 1.00                                               & \textbf{1.00}                                     & 1.00                                               & 1.00                                               & 1.00                                               & \textbf{1.00}                                     \\
F-measure                                         & 0.97                                               & 0.97                                               & 0.95                                               & \textbf{0.96}                                     & 0.68                                               & 0.77                                               & 0.74                                               & \textbf{0.73}                                     & 0.60                                               & 0.64                                               & 0.60                                               & \textbf{0.61}   \\
Accuracy(\%)                                          & 96.54                                              & 96.47                                              & 94.99                                              & \textbf{96.00}                                    & 67.59                                              & 76.96                                              & 73.47                                              & \textbf{72.67}                                    & 60.45                                              & 64.47                                              & 60.00                                              & \textbf{61.64}                                    

\end{tabular}
}
\caption{Results of single-modality face identification and speaker identification systems with three different feature sets for three folds and presentation of average values}
\label{tab:t1}

\small
\centering
\adjustbox{width=15cm}{
\begin{tabular}{
>{\columncolor[HTML]{FFFFFF}}l |
>{\columncolor[HTML]{FFFFFF}}c
>{\columncolor[HTML]{FFFFFF}}c
>{\columncolor[HTML]{FFFFFF}}c
>{\columncolor[HTML]{FFFFFF}}c |
>{\columncolor[HTML]{FFFFFF}}c
>{\columncolor[HTML]{FFFFFF}}c
>{\columncolor[HTML]{FFFFFF}}c
>{\columncolor[HTML]{FFFFFF}}c |
>{\columncolor[HTML]{FFFFFF}}c
>{\columncolor[HTML]{FFFFFF}}c
>{\columncolor[HTML]{FFFFFF}}c
>{\columncolor[HTML]{FFFFFF}}c |}
\cellcolor[HTML]{FFFFFF}                          & \multicolumn{4}{c|}{\cellcolor[HTML]{D9D9D9}Sensor Fusion}                                                                                                                                                       & \multicolumn{4}{c|}{\cellcolor[HTML]{BFBFBF}Score Fusion}                                                                                                                                                        & \multicolumn{4}{c|}{\cellcolor[HTML]{A6A6A6}Feature Fusion}                                                                                                                                                      \\ \cline{2-13}
\multirow{-2}{*}{\cellcolor[HTML]{FFFFFF}Metrics} & \multicolumn{1}{l}{\cellcolor[HTML]{FFFFFF}Fold 1} & \multicolumn{1}{l}{\cellcolor[HTML]{FFFFFF}Fold 2} & \multicolumn{1}{l}{\cellcolor[HTML]{FFFFFF}Fold 3} & \multicolumn{1}{l|}{\cellcolor[HTML]{D9D9D9}Avg.} & \multicolumn{1}{l}{\cellcolor[HTML]{FFFFFF}Fold 1} & \multicolumn{1}{l}{\cellcolor[HTML]{FFFFFF}Fold 2} & \multicolumn{1}{l}{\cellcolor[HTML]{FFFFFF}Fold 3} & \multicolumn{1}{l|}{\cellcolor[HTML]{BFBFBF}Avg.} & \multicolumn{1}{l}{\cellcolor[HTML]{FFFFFF}Fold 1} & \multicolumn{1}{l}{\cellcolor[HTML]{FFFFFF}Fold 2} & \multicolumn{1}{l}{\cellcolor[HTML]{FFFFFF}Fold 3} & \multicolumn{1}{l|}{\cellcolor[HTML]{A6A6A6}Avg.} \\ \hline
Precision                                         & 0.95                                               & 0.95                                               & 0.91                                               & \textbf{0.94}                                     & 0.97                                               & 0.97                                               & 0.95                                               & \textbf{0.96}                                     & 0.99                                               & 0.99                                               & 0.98                                               & \textbf{0.99}                                     \\
Sensitivity                                       & 0.95                                               & 0.95                                               & 0.91                                               & \textbf{0.94}                                     & 0.97                                               & 0.97                                               & 0.95                                               & \textbf{0.96}                                     & 0.99                                               & 0.99                                               & 0.98                                               & \textbf{0.99}                                     \\
Specificity                                       & 1.00                                               & 1.00                                               & 1.00                                               & \textbf{1.00}                                     & 1.00                                               & 1.00                                               & 1.00                                               & \textbf{1.00}                                     & 1.00                                               & 1.00                                               & 1.00                                               & \textbf{1.00}                                     \\
F-measure                                         & 0.95                                               & 0.95                                               & 0.91                                               & \textbf{0.94}                                     & 0.97                                               & 0.97                                               & 0.95                                               & \textbf{0.96}                                     & 0.99                                               & 0.99                                               & 0.98                                               & \textbf{0.99}  \\
Accuracy(\%)                                          & 95.02                                              & 95.10                                              & 90.70                                              & \textbf{93.61}                                    & 96.73                                              & 96.81                                              & 95.18                                              & \textbf{96.24}                                    & 98.59                                              & 98.94                                              & 97.59                                              & \textbf{98.37}                                    

\end{tabular}
}
\caption{Performance of proposed multimodal identification systems in three different fusion strategies based on gammatonegram and facial features}
\label{tab:t2}
%

\small
\centering
\adjustbox{width=11cm}{
\begin{tabular}{
>{\columncolor[HTML]{FFFFFF}}l |
>{\columncolor[HTML]{FFFFFF}}c
>{\columncolor[HTML]{FFFFFF}}c
>{\columncolor[HTML]{FFFFFF}}c
>{\columncolor[HTML]{FFFFFF}}c |
>{\columncolor[HTML]{FFFFFF}}c
>{\columncolor[HTML]{FFFFFF}}c
>{\columncolor[HTML]{FFFFFF}}c
>{\columncolor[HTML]{FFFFFF}}c |}
\cellcolor[HTML]{FFFFFF}                          & \multicolumn{4}{c|}{\cellcolor[HTML]{D0CECE}Score Fusion}                                                                                                                                                    & \multicolumn{4}{c|}{\cellcolor[HTML]{AEAAAA}Feature Fusion}                                                                                                                                                 \\ \cline{2-9}
\multirow{-2}{*}{\cellcolor[HTML]{FFFFFF}Metrics} & \multicolumn{1}{l}{\cellcolor[HTML]{FFFFFF}Fold 1} & \multicolumn{1}{l}{\cellcolor[HTML]{FFFFFF}Fold 2} & \multicolumn{1}{l}{\cellcolor[HTML]{FFFFFF}Fold 3} & \multicolumn{1}{l|}{\cellcolor[HTML]{D0CECE}Avg.} & \multicolumn{1}{l}{\cellcolor[HTML]{FFFFFF}Fold 1} & \multicolumn{1}{l}{\cellcolor[HTML]{FFFFFF}Fold 2} & \multicolumn{1}{l}{\cellcolor[HTML]{FFFFFF}Fold 3} & \multicolumn{1}{l|}{\cellcolor[HTML]{AEAAAA}Avg.} \\ \hline
Precision                                         & 0.97                                               & 0.97                                               & 0.96                                               & \textbf{0.97}                                     & 0.99                                               & 0.99                                               & 0.97                                               & \textbf{0.98}                                     \\
Sensitivity                       & 0.97                                               & 0.97                                               & 0.96                                               & \textbf{0.97}                                     & 0.99                                               & 0.99                                               & 0.97                                               & \textbf{0.98}                                     \\
Specificity                                        & 1.00                                               & 1.00                                               & 1.00                                               & \textbf{1.00}                                     & 1.00                                               & 1.00                                               & 1.00                                               & \textbf{1.00}                                     \\
F-measure                                         & 0.97                                               & 0.97                                               & 0.96                                               & \textbf{0.97}                                     & 0.99                                               & 0.99                                               & 0.97                                               & \textbf{0.98}  \\
Accuracy(\%)                                            & 96.88                                              & 97.27                                              & 95.94                                              & \textbf{96.70}                                    & 98.81                                              & 98.88                                              & 97.28                                              & \textbf{98.33}                                   

\end{tabular}
}
\caption{Results of proposed multimodal identification systems in two different fusion strategies based on x-Vector and facial features}
\label{tab:t3}
\end{table*}
\subsection{Person Identification}
\label{ssec:idntidnt}
According to the results in Table \ref{tab:t1}, the performance of the proposed person identification systems in the single modality can be observed in face identification and two other speaker identification separated based on the x-vector and gammatonegram representation methodologies. The table includes the achievements for each fold separately and the average performance. Based on the results, it is evident that the system performs better in face single modality recognition than voice. The accuracy achieved by the FaceNet is 96.00\%, while the VoiceNet reached an accuracy of 72.67\% using x-vector and an accuracy of 61.64\% based on gammatonegram. It seems that the x-vector could represent the speaker feature better than the gammatonegram in single modality mode.

Other parameters that provide more insight into the systems' performance can also be found in Table \ref{tab:t1}. These parameters present the ability of systems to accept correct utterances and reject incorrect utterances for each class. It is worth noting that the presence of babble noise and low sound quality significantly impact the performance of VoiceNets, resulting in its lower accuracy compared to FaceNet.

In addition, it can be seen that the modalities can complement each other while being independent of each other. Table \ref{tab:t2} presents the results of three multimodal strategies, each based on different fusion methods. In this table, gammatonegram is used for voice and FaceNet features for face modality. The first part of the table illustrates the performance of the proposed multimodal learning system in sensor fusion mode. Despite achieving precision, sensitivity, specificity, F-measure, and accuracy percentage of 0.94, 0.94, 1, 0.94, and 93.61\%, respectively, the multimodal system in sensor fusion mode underperforms compared to the FaceNet single modality system.

This suggests that gammatonegram as a speech presentation method can potentially confuse the sensor-level multimodal system, leading to a decrease in performance in comparison with the single modality system. Although the multimodal learning system with input-level fusion of image and speech data does not improve efficiency, it provides valuable insights to researchers. The findings indicate that the fusion of sensors for these two modalities may not significantly enhance performance, underscoring the need for careful consideration of fusion strategies and modality compatibility in multimodal learning systems.

Based on the investigation in Table \ref{tab:t2}, the multimodal system in score fusion mode has been evaluated using the information from previous single-modality systems based on facial features and gammatonegram representation. Each network's final softmax layer data is used as input for the aggregated softmax layer in multimodal learning. The average performance across three different folds shows that the system achieves better precision, sensitivity, specificity, F-measure, and accuracy scores than the single modality mode. Specifically, the system achieves 96.24\% accuracy, which is 0.24\% higher than the result obtained by the FaceNet single-modality system.

Table \ref{tab:t2} reveals that the proposed softmax layer in feature fusion mode demonstrates high efficiency, achieving 98.37\% accuracy, 0.99 precision, 0.99 sensitivity, 1 specificity, and 0.99 F-measure. These results signify a significant improvement in the accuracy of the person identification task, with a 2.37\% increase compared to the performance of FaceNet in single modality mode. This underscores the effectiveness of the feature fusion strategy in enhancing the system's accuracy. The additional source data from different modalities provide extra information for identification, as they express different aspects of the same class. For instance, in fold 1, the speaker identification has the lowest performance compared to the other two folds, but the facial recognition for this fold is done with high accuracy. This demonstrates that modality fusion can compensate for the shortcomings of single modalities.
\setlength{\tabcolsep}{6pt} 
\renewcommand{\arraystretch}{1.2} 

\begin{table}
\small
\centering
\adjustbox{width=7.5cm}{
\begin{tabular}{
>{\columncolor[HTML]{FFFFFF}}l |
>{\columncolor[HTML]{FFFFFF}}c
>{\columncolor[HTML]{FFFFFF}}c
>{\columncolor[HTML]{FFFFFF}}c
>{\columncolor[HTML]{FFFFFF}}c |
>{\columncolor[HTML]{FFFFFF}}c
>{\columncolor[HTML]{FFFFFF}}c
>{\columncolor[HTML]{FFFFFF}}c
>{\columncolor[HTML]{FFFFFF}}c |}
\cellcolor[HTML]{FFFFFF}                          & \multicolumn{4}{c|}{\cellcolor[HTML]{D0CECE}Face Verification}                                                                                                                                                    & \multicolumn{4}{c|}{\cellcolor[HTML]{AEAAAA}Speaker Verification}                                                                                                                                                 \\ \cline{2-9}
\multirow{-2}{*}{\cellcolor[HTML]{FFFFFF}Metrics} & \multicolumn{1}{l}{\cellcolor[HTML]{FFFFFF}Fold 1} & \multicolumn{1}{l}{\cellcolor[HTML]{FFFFFF}Fold 2} & \multicolumn{1}{l}{\cellcolor[HTML]{FFFFFF}Fold 3} & \multicolumn{1}{l|}{\cellcolor[HTML]{D0CECE}Avg.} & \multicolumn{1}{l}{\cellcolor[HTML]{FFFFFF}Fold 1} & \multicolumn{1}{l}{\cellcolor[HTML]{FFFFFF}Fold 2} & \multicolumn{1}{l}{\cellcolor[HTML]{FFFFFF}Fold 3} & \multicolumn{1}{l|}{\cellcolor[HTML]{AEAAAA}Avg.} \\ \hline
EER(\%)                                          & 0.48                                               & 0.50                                               & 2.07                                               & \textbf{1.01}                                     & 5.99                                               & 5.37                                               & 3.65                                               & \textbf{5.12}                \\
\end{tabular}
}
\caption{Results of proposed person verification systems in two single modes}
\label{tab:t4}

\small
\centering
\adjustbox{width=7.5cm}{
\begin{tabular}{
>{\columncolor[HTML]{FFFFFF}}l |
>{\columncolor[HTML]{FFFFFF}}c
>{\columncolor[HTML]{FFFFFF}}c
>{\columncolor[HTML]{FFFFFF}}c
>{\columncolor[HTML]{FFFFFF}}c |
>{\columncolor[HTML]{FFFFFF}}c
>{\columncolor[HTML]{FFFFFF}}c
>{\columncolor[HTML]{FFFFFF}}c
>{\columncolor[HTML]{FFFFFF}}c |}
\cellcolor[HTML]{FFFFFF}                          & \multicolumn{4}{c|}{\cellcolor[HTML]{D0CECE}Gammatonegram and FaceNet}                                                                                                                                                    & \multicolumn{4}{c|}{\cellcolor[HTML]{AEAAAA}x-Vector and FaceNet}                                                                                                                                                 \\ \cline{2-9}
\multirow{-2}{*}{\cellcolor[HTML]{FFFFFF}Metrics} & \multicolumn{1}{l}{\cellcolor[HTML]{FFFFFF}Fold 1} & \multicolumn{1}{l}{\cellcolor[HTML]{FFFFFF}Fold 2} & \multicolumn{1}{l}{\cellcolor[HTML]{FFFFFF}Fold 3} & \multicolumn{1}{l|}{\cellcolor[HTML]{D0CECE}Avg.} & \multicolumn{1}{l}{\cellcolor[HTML]{FFFFFF}Fold 1} & \multicolumn{1}{l}{\cellcolor[HTML]{FFFFFF}Fold 2} & \multicolumn{1}{l}{\cellcolor[HTML]{FFFFFF}Fold 3} & \multicolumn{1}{l|}{\cellcolor[HTML]{AEAAAA}Avg.} \\ \hline
EER(\%)                                          & 0.44                                               & 0.41                                               & 1.61                                               & \textbf{0.82}                                     & 0.37                                               & 0.36                                               & 1.15                                               & \textbf{0.62}                \\
\end{tabular}
}
\caption{Results of proposed person verification systems in feature fusion modes based on two different speech features}
\label{tab:t5}
\end{table}
Based on Table \ref{tab:t3}, the proposed softmax layer for facial features and x-vector in score fusion mode reaches 96.7\% accuracy on average, which is better than the single modality mode. However, the feature fusion mode showcases remarkable effectiveness, achieving an accuracy of 98.33\%, precision and sensitivity scores of 0.98 each, a specificity of 1, and an F-measure of 0.98. These findings underscore the successful enhancement in accuracy for the person identification task by 2.33\% compared to FaceNet as the best achievement in single modality mode.

The average achievements of all five modes show that FaceNet performed the best in single-modality scenarios. Despite VoiceNet's lower efficiency, when combined with the face embedding feature using gammatonegram and x-vector, it improved the identification system's performance in two scenarios: feature fusion and score fusion. Our best achievement was in the feature fusion mode, where we combined facial features with the gammatonegram.
\subsection{Person Verification}
\label{ssec:idntver}
In this section, we investigate the results of the proposed system for the person verification task. Based on the results obtained in person identification, the feature-level fusion has shown the best performance. Therefore, in the verification scenario, we only evaluate the fusion of features consisting of gammatonegram, x-vector, and FaceNet's activations in multimodal mode, besides two typical single modality-based person verification.
In this feature fusion mode, the proposed person verification system utilizes Within-Class Covariance Normalization (WCCN) and Linear Discriminant Analysis (LDA) to reduce intra-class variation and decrease the dimension of the feature vector, respectively. The LDA output yields a vector with 150 components as an eigenvector from the original vector with 1,024 elements.

Furthermore, Gaussian Probabilistic Linear Discriminant Analysis (GPLDA) is employed for decision-making. It is trained with 20 iterations and uses vectors with a length of 150 components. These parameters remain fixed in both single and multimodal scenarios. The results of this evaluation can be found in Table \ref{tab:t4}.

In the single modality mode in verification tasks, both proposed FaceNet and VoiceNet are used as feature extractors, and their features are fed into GPLDA for decision-making in a two-class classification. The results indicate that the average EER in the speaker verification using voice modality is 5.12\%, while in the face verification mode, it is 1.01\%. As expected, the system based on face modality performs better due to the superior performance of the proposed face feature extraction. In multimodal mode, we designed two systems based on two different voice features.

In the fusion of x-vector and facial features, the system achieves an average EER of 0.62\% in the feature fusion mode. however, in concatenating the gammatonegram feature vector with the facial feature vector, the system reaches 0.82\% of EER. Similar to the speaker identification scenario, this demonstrates an improved performance compared to the single modality mode.
\section{Discussion}
\label{sec:dis}
visualization of the activity status of layers can also provide useful information for understanding CNN's functionality. Usually, an appropriate cognition of what happens inside the CNN network gets little attention, and one of the approaches is to consider the CNN as a black box. However, there are helpful visualization methods to understand what is happening inside of a CNN. We depict a representation method called the Locally-Interpretable Model-agnostic Explanation (LIME) technique \cite{ribeiro2016should} to show the crucial parts of the image that play a more significant role in network decision-making. 

\begin{figure}[ht]
  \centering
  \begin{minipage}[b]{0.38\textwidth}
    \centering
    \includegraphics[width=\textwidth]{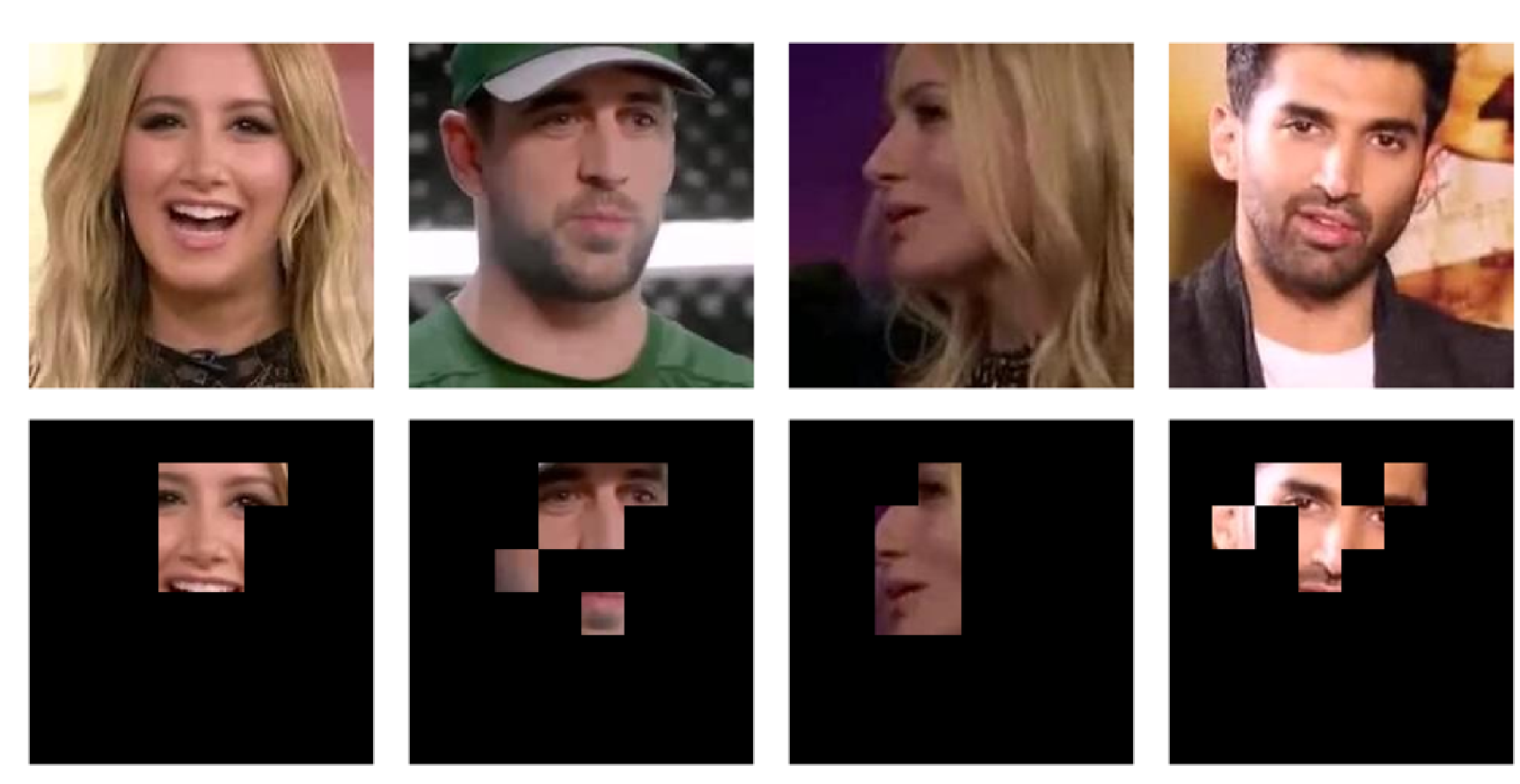} 
    \caption{Crucial parts of face images for decision-making in the FaceNet}
    \label{fig:f2}
  \end{minipage}
  \hfill 
  \begin{minipage}[b]{0.38\textwidth}
    \centering
    \includegraphics[width=\textwidth]{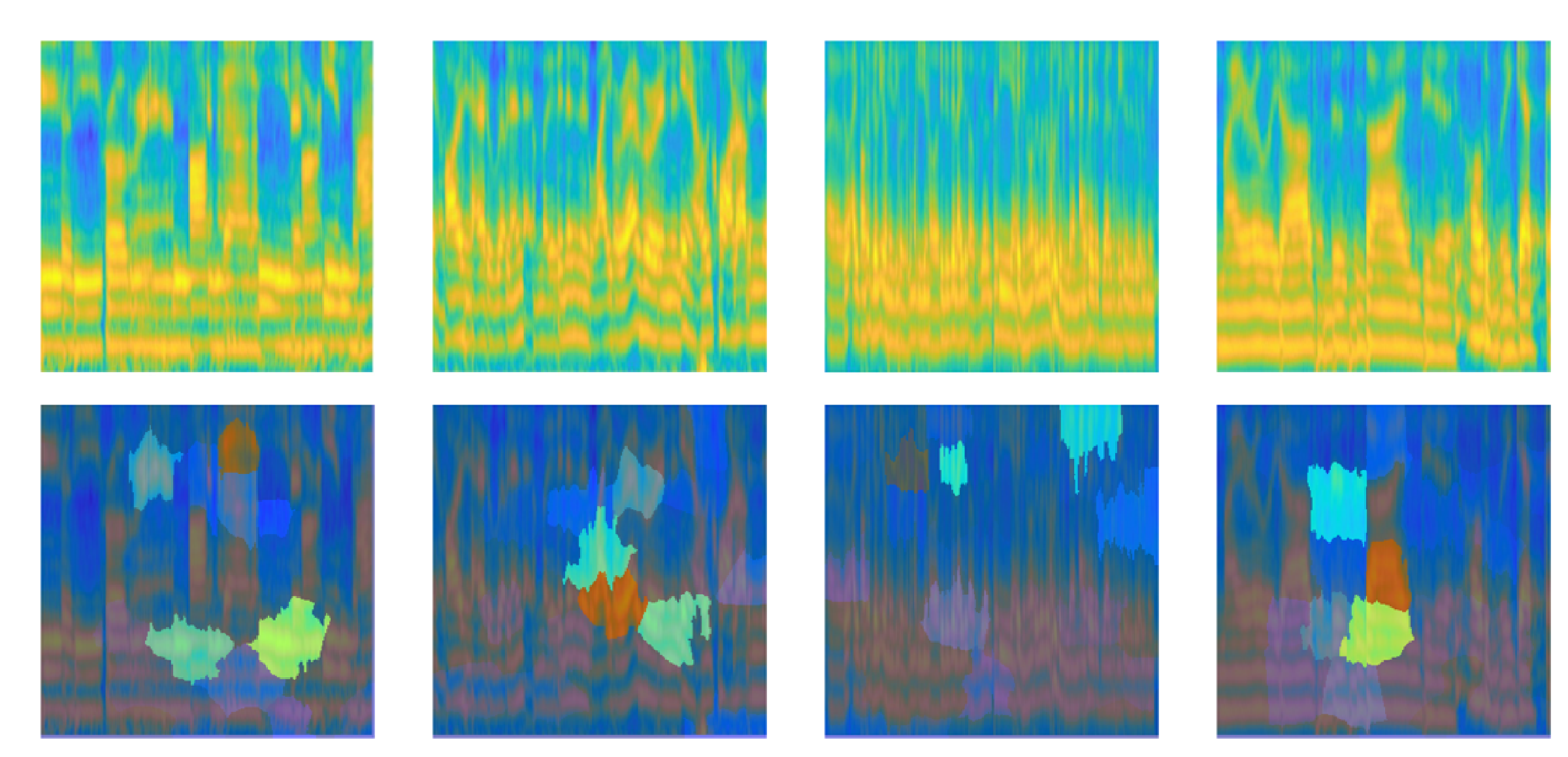} 
    \caption{Important part of gammatonegram image from the viewpoint of Darknet19 network for classification}
    \label{fig:f3}
  \end{minipage}
\end{figure}

This visualization was made using 2D-CNNs consisting of FaceNet in Figure \ref{fig:f2} and gammatonegram-based speaker identifiers in Figure \ref{fig:f3}.
These figures make it possible to understand which parts of the images are focused on by the network. It is also possible to understand whether the network is focused on the essential and discriminative parts of the image or not. The image segmentation colors show each part's priority according to a color bar. By examining the map and its corresponding image, it can be seen that the network focused on the distinguishing parts of the image, which is the reason for the acceptable result for FaceNet and weak performance on speaker identification using the gammatonegram feature.

In this work, we tried to analyze different strategies for audio-visual modality fusion in the identity recognition task. However, it could be helpful to compare the results of the proposed systems with previous work to understand the scale of the metrics. The results of the present work depict that utilizing gammatonegram representation for voice modality and VGGFace2 pre-trained network for face modality can properly depict the identity information in the identification task. However, utilizing x-vector as a voice feature vector beside fine-tuned VGGFace2 in the verification tasks can present the best result. 

\section{Conclusion}
\label{sec:cocl}
In this work, we analyzed different strategies for audio-visual modality fusion in the identity recognition tasks. The results of the present work depict that utilizing gammatonegram representation for voice modality and VGGFace2 pre-trained network for face modality can properly depict the identity information. 

 This study developed two separate single-modality systems for voice and face in two identification and verification tasks. The FaceNet architecture utilized VGGFace2, while we had two VoiceNet, the first one was trained from scratch to extract the x-vector, and the second one was based on gammatonegram representation and fine-tuning of pre-trained Darknet19. Based on these systems, multimodal learning was applied in three fusion modes, as described in the article. A softmax layer was the classifier in feature fusion and score fusion approaches in the person identification task. However, for person verification, LDA was used for dimension reduction, and GPLDA was utilized for decision-making. The evaluation was conducted on 118 speakers from the VoxCeleb2 dataset. The results demonstrated that combining speech and face modalities using multimodal learning outperformed the single-mode approach in both identification and verification tasks. Additionally, the feature fusion mode was found to be the most effective strategy for these two modalities.
 
Future studies could enhance this research by investigating more efficient speech features, such as deep belief networks in autoencoder architecture, to be used as input for the proposed VoiceNet for x-vector extraction. Moreover, utilizing para-linguistic systems, such as gender recognition from face or voice, could offer valuable insights in a score fusion scenario.




\bibliographystyle{IEEEtran}
\bibliography{Template2.bib}

\end{document}